\documentclass[12pt]{article}
\usepackage{amsfonts}

\def\beq{\begin{equation}}
\def\eeq{\end{equation}}
\def\bea{\begin{eqnarray}}
\def\eea{\end{eqnarray}}
\def\ba{\begin{array}}
\def\ea{\end{array}}
\def\pa{\partial}

\def\nn{\nonumber}

\def\Tr{{\rm Tr}}

\def\ga{\gamma}
\def\Ga{\Gamma}

\def\de{\delta}

\def\al{\alpha}

\def\om{\omega}
\def\la{\lambda}
\def\La{\Lambda}
\def\eps{\epsilon}
\def\phc{\phi_c}

\def\rd{{\rm d}}
\def\GeV{{\rm GeV}}

\def\rd{{\rm d}}
\def\ri{{\rm i}}

\begin{document}
\hfill AEI-2007-146\\

\centerline{\bf\Large Effective Action, Conformal Anomaly and}
\vspace{0.2cm}
\centerline{\bf\Large the Issue of Quadratic Divergences}

\vspace{5mm}
\centerline{\bf Krzysztof A. Meissner${}^{1}$ and Hermann Nicolai${}^2$}

\vspace{5mm}
\begin{center}
{\it ${}^1$ Institute of Theoretical Physics, University of Warsaw,\\
Ho\.za 69, 00-681 Warsaw, Poland\\ ${}^2$ Max-Planck-Institut
f\"ur Gravitationsphysik (Albert-Einstein-Institut),\\
M\"uhlenberg 1, D-14476 Potsdam, Germany}
\end{center}

\begin{abstract}
\footnotesize{For massless $\phi^4$ theory, we explicitly compute the
lowest order non-local contributions to the one-loop effective
action required for the determination of the trace anomaly.
Imposing {\em exact} conformal invariance of the {\em local} part
of the effective action, we argue that the issue of quadratic
divergences does not arise in a theory where exact conformal
symmetry is only broken by quantum effects. Conformal symmetry can
thus replace low energy supersymmetry as a possible guide towards
stabilizing the weak scale and solving the hierarchy problem, if
$(i)$ there are no intermediate scales between the weak scale and the
Planck scale, and $(ii)$ the running couplings exhibit neither Landau 
poles nor instabilities over this whole range of energies.}
\end{abstract}

\section{Introduction}

The present work is motivated  by the possibility that the standard
model (SM) of elementary particle physics could arise from an exactly
conformal invariant theory~\footnote{Classic references on conformal
 symmetry in (quantum) field theory are \cite{MS,CCJ}.}. All observed 
mass scales of particle physics and their smallness {\it vis-\`a-vis}
the Planck scale might thus be explained via the quantum mechanical
breaking of conformal symmetry induced by the explicit breaking of conformal
symmetry that necessarily accompanies any regularization. The explanation
of small mass scales via a conformal anomaly would thus be natural in
the sense of \cite{tH}, that is, in terms of a `nearby' exact conformal
symmetry.

Although the attractiveness of such a scenario has been appreciated
for a long time since the seminal work of Coleman and Weinberg \cite{CW},
it appears that attempts at its concrete implementation have so far
met only with limited success. In part this may be due to a widely
held expectation, according to which the existence of large intermediate
scales between the weak scale and the Planck scale is unavoidable
if one wants to understand physics beyond the SM, as exemplified
by grand unification (GUTs) and the conventional explanation of small
neutrino masses via the seesaw mechanism (both of which require
new scales above $10^{11} \, \GeV$). In this view, low energy
physics would be separated from the Planck scale by a cascade of
new scales, and there would be no place for conformal invariance,
even if only approximate.

In a recent paper \cite{MN} we have advocated a different scenario, which
is based on the assumption of an exactly conformally invariant tree level
Lagrangian, and the absence of new large scales below the Planck scale --
a scenario often referred to as the `grand desert'~\footnote{See also
 \cite{ST,Shap} and references therein for a related, but different
 `grand desert' scenario where conformal invariance is broken by
 explicit mass terms for the scalar fields. A similar model with
 extra scalar fields, but without right-chiral neutrinos is also
 considered in \cite{EQ}.}. Evidence was presented
that, with the incorporation of massive neutrinos, all observed
features of the SM can be reproduced in this way. Apart from the
inclusion of neutrinos, a crucial difference with earlier attempts
in this direction (see e.g.
\cite{CW1}) was the imposition of an extra consistency postulate,
namely the requirement that the evolution of all couplings
according to the renormalization group equations should be such
that the theory remains viable up to a very large scale, that is,
there should exist neither Landau poles nor instabilities below
that scale.~\footnote{Landau poles exist {\em for all} extensions
of the SM,
 so that none of these extensions is likely to exist as a rigorous
 quantum field theory (see e.g. \cite{GJ,R}). At issue is only the
 {\em scale} at which these poles occur.} This obviously requires
a special conspiracy of the SM parameters, but as shown in
\cite{MN} there does exist a `window' compatible with known
phenomenology for which the model can survive up to the Planck
scale, and which predicts the Higgs mass(es) to be above $200\,
\GeV$ .

In this article we wish to address a very specific technical
issue, namely the computation of the effective action at one loop
beyond the constant field approximation underlying the one-loop
effective potential of \cite{CW}, and the incorporation of {\em
non-local} corrections. For simplicity we will restrict attention
to scalar fields, and only consider the massless (and thus
classically conformally invariant) $\phi^4$ theory. Although there
is a huge literature on trace anomalies in gauge theories (see
e.g. \cite{CDJ,F}) and in curved backgrounds (see e.g.\cite{F,Duff}),
it appears that the non-local effective action for a self-interacting
scalar field has not received so much attention.
Because one cannot hope to derive an analytical expression for the
complete one-loop effective action, various approximation schemes have
been devised in previous work. For instance, in \cite{BO} a `quasi-local'
approximation is developed, while \cite{W} sets up a formalism based
on `average actions'; more recently, ref.~\cite{BM} derives expressions
for the non-local effective action in the small and large field limits,
respectively. By contrast, we here focus on the {\em conformal
properties} in that we aim for determining only that part of the
one-loop effective action which captures the anomalous behavior
under conformal transformations (see also \cite{BOS}). In this way,
we can show explicitly that the conformal anomaly [cf. eq.~(\ref{anom})]
is given by a {\em local} expression which itself arises as the
variation of a non-local functional [cf. eq.~(\ref{Seff1})]. That
the anomaly {\em must} be local follows from general arguments,
and constitutes a crucial consistency check on any approximation
scheme for the non-local effective action.

Our main purpose in taking up again these (rather old) questions
here is to reconsider their implications for physics beyond the
standard model, in particular the question of naturalness and
stability of the weak scale. The appearance of quadratic
divergences in theories with scalars is usually invoked as the
main (theoretical) argument for low energy supersymmetry. By
contrast, we here argue -- following earlier arguments by Bardeen
\cite{Bardeen} -- that (quantum mechanically broken) conformal
symmetry may provide an equally good mechanism for the stabilization
of the weak scale, if we can impose {\em exact} conformal invariance
on the counterterm dependent terms of the effective action (for a
discussion of scale stability with softly broken conformal symmetry
see also \cite{FV}, which likewise casts some doubt on the usual
arguments for low energy supersymmetry). The main role of supersymmetry
could then be in ensuring finiteness of quantum gravity.

The plan of this paper is as follows. In section~2 we review the
construction of the one-loop effective action for pure $\phi^4$
theory; the conformal properties of the resulting expression are
discussed in section~3. In section~4 we consider the implications
of these results for the issue of quadratic divergences in scalar
field theory. Some useful formulae are collected in the Appendix.

\section{Massless $\phi^4$ theory revisited}

The simplest (non-trivial) example of a classically conformally
invariant theory in four space-time dimensions is massless $\phi^4$
theory, whose action reads~\footnote{For convenience, we will work
  with a Euclidean metric throughout this paper.}
\beq\label{S}
S^E_{cl}=\int\rd^4 x \sqrt{g} \left(- \frac12 \phi \Box_g \phi
      + \frac1{12} R\phi^2 + \frac{\la}{4}\phi^4\right)
\label{scl}
\eeq
where
\beq\label{LB}
\Box_g := \frac1{\sqrt{g}} \partial_\mu (\sqrt{g} g^{\mu\nu} \partial_\nu)
\eeq
is the generally covariant Laplace-Beltrami operator. It is a standard
result \cite{CCJ} that the addition of the term involving the Ricci scalar
$R$ in (\ref{S}) makes the action also invariant under local re-scalings
(Weyl transformations). Our main concern in this paper is the flat space
theory, but we nevertheless have included the Weyl invariant gravitational
couplings here because they will contribute to the conformal (trace)
anomaly even in the flat space limit. Accordingly, we set
$g_{\mu\nu}(x) = \eta_{\mu\nu}$, except in those places relevant to
the computation of the trace anomaly.

As is well known \cite{Z,IZ,Weinberg}, the effective action is defined
as the Legendre transform
\beq
\Ga[\phc] = W[J] - \int d^4x J(x) \phc (x)
\eeq
with the generating functional $W[J]$ of one-particle irreducible (1PI)
Green's functions, where the classical field $\phc(x)$ is defined by
\beq
\phc (x) = \frac{\de W[J]}{\de J(x)}
\eeq
The effective action can be expanded in powers of $\hbar$ (``loop expansion'')
\beq
\Ga[\phc]=\Ga^{(0)}[\phc]+\hbar\Ga^{(1)}[\phc]  +  \ldots
\eeq
where
\beq
\Ga^{(0)}[\phc]=S^E_{cl}(\phc)
\eeq
It is also well known that the one-loop contribution to the
effective action is given by \cite{IZ}
\beq
\Ga^{(1)}[\phc]=\frac12 \Tr\ln\frac{\de^2 S_{cl}}{\de\phi(x)\de\phi (y)}
\eeq
For the classical action (\ref{scl}) one thus straightforwardly
derives the (still formal) result
\beq
\Ga^{(1)}[\phc]=\frac12 \Tr\ln\left(-\Box+3\la\phc^2(x)\right)
\label{Seff0}
\eeq
which, in turn, can be rewritten as (dropping an infinite constant)
\beq
\Ga^{(1)}[\phc]=-\frac12 \Tr\ln\left(1-\frac{3\la\phc^2(x)}
{-\Box+3\la\phc^2(x)}\right)
\label{Seffdif}
\eeq
This expression can be (formally) expanded as
\bea
\Ga^{(1)}[\phc] &=& \frac{3\la}2 \int d^4x \sqrt{g(x)}
    \, D(x,x;\phc) \,\phc^2(x) \; +
    \label{gaform}\\
    && \!\!\!\!\!\!\!\!\!\!\!\!\!\!\!\!\!\!\!\!\!
   + \; \frac{9\la^2}4\int d^4x \sqrt{g(x)}
   \int d^4y \sqrt{g(y)}\, D(x,y;\phc) \, \phc^2(y)\,
        D(y,x;\phc) \, \phc^2(x) \; + \dots  \nn
\eea
where the field dependent propagator $D(x,y;\phc)$ is defined as
\beq
\big(-\Box_g + \frac16 R + 3\la \phc^2(x)\big) D_g(x,y;\phc)
= g^{-3/8}(x) \delta^{(4)}(x,y) g^{-1/8}(y)
\label{propag}
\eeq
The gravitational couplings have been re-instated here so we can
discuss the conformal properties of these objects. Namely, under
local conformal transformations we have
\beq
g_{\mu\nu}(x)\to e^{-2\om(x)}g_{\mu\nu}(x)\; ,\ \ \
\phc(x)\to e^{\om(x)}\phc(x)
\label{conftr}
\eeq
The conformally covariant Laplacian transforms as
\beq
\left(-\Box_g +\frac16 R\right)\phc(x)\to e^{3\om(x)}\left(-\Box_g
 +\frac16 R\right)\phc(x)
\label{confinv}
\eeq
It is important here that the operator in parentheses acts on a
quantity of conformal weight one; it would not be conformally
covariant when acting on a field of different conformal weight,
such as $\phc^2(x)$. From these formulae we deduce the
transformation properties of the propagator (\ref{propag}) under
conformal re-scalings, {\it viz.}
\beq
D(x,y;\phc) \; \longrightarrow \; e^{\om(x)} D(x,y;\phc) \, e^{\om (y)}
\eeq
Therefore (\ref{gaform}) is not only generally covariant, but in
addition also formally invariant under conformal transformations ---
only formally, because the first two terms in the expansion are
divergent and require renormalization which introduces, as we will
see, a conformal anomaly.

Let us start with a constant field $\phc(x)=\phi_0$. The
propagator reads then
\beq
D(x-y;M)\equiv D(x,y;\phc) \big|_{\phc(x)=\phi_0} =
\int\frac{\rd^4 k}{(2\pi)^4}\frac{e^{-\ri k(x-y)}}{k^2+ M^2}
\eeq
where $M^2=3\la\phi_0^2$. Summing up the series we obtain the standard
result
\beq
\Ga^{(1)}[\phi_0]=\frac12\int\rd^4 x \int\frac{\rd^4 k}{(2\pi)^4}
\ln\left(1+\frac{M^2}{k^2}\right)
\label{CWa}
\eeq
Before we proceed to the general case we emphasize the following point.
To renormalize the theory, we should impose {\em exact conformal
invariance on the counterterm dependent terms of the effective
action}. This requirement in particular {\em leaves no room for
a finite mass term} in the local part of (\ref{gaform}). For
the further computation it is therefore more convenient to work with
dimensional regularization (see e.g.
\cite{IZ,Weinberg}) because the latter automatically satisfies the
requirement of not introducing any local terms that break
conformal invariance in the effective action at any given order.
In particular, all divergent counterterms (poles in $\eps$) are
conformally invariant. The regularization is performed in the
usual way by continuing all loop integrals to $D=4-2\eps$ dimensions
via the replacement
\beq
\int \frac{d^4 k}{(2\pi)^4} \;\; \longrightarrow \;\;
(C v^2)^\epsilon
\int \frac{d^{4-2\epsilon} k}{(2\pi)^{4-2\epsilon}}
\eeq
where $C=e^\ga/(4\pi)$. Like any other regulator, this procedure
breaks classical conformal invariance, here via the dimensionful
parameter $v$. Using the formulae from the Appendix we get, in the
constant field case,
\beq
\Ga^{(1)}[\phi_0]=\frac{9\la^2}{64\pi^2}\int\rd^4 x\, \phi_0^4
\left[-\frac{1}{\eps}+
\left(\ln\frac{3\la\phi_0^2}{v^2}-\frac32\right)\right]
\; + \; {\cal{O}} (\eps)
\label{CWres}
\eeq
which, after renormalization, is just the familiar Coleman-Weinberg
effective potential \cite{CW}.

When the field $\phi_c(x)$ depends on $x$ we obviously cannot sum
the series in (\ref{gaform}). Since we are mainly interested in
the conformal anomaly we analyze only the first two terms of the
expansion (those displayed explicitly in (\ref{gaform})) because all other
terms are convergent and therefore
conformally invariant. Let us write these two terms expanding the
propagators around a constant field $\phi_0^2$
\bea
\Ga^{(1)}(\phc)&=&
\frac{3\la}{2}
\int\rd^4 x\, \phc^2(x)\int\frac{\rd^4 k}{(2\pi)^4}
\frac{1}{k^2+3\la\phi_0^2} \;  \label{gaexp}\\
&& \!\!\!\!\!\!\!\!\!\!\!\!\!\!\!\!\!\!\!\!\!\!\!
 -\; \frac{9\la^2}{32\pi^2}\int\rd^4 x\rd^4 y\,
(\phc^2(x)-\phi_0^2)\, K(x-y;3\la\phi_0^2) \, \phc^2(y) \nn\\
&&
\!\!\!\!\!\!\!\!\!\!\!\!\!\!\!\!\!\!\!\!\!\!\!
 +\; \frac{9\la^2}{64\pi^2}\int\rd^4 x\rd^4 y\,
\phc^2(x)\, K(x-y;3\la\phi_0^2) \, \phc^2(y)+\ldots
\nn
\eea
where the dots stand for higher powers of $(\phc^2(x)-\phi_0^2)$
and higher number of propagators (so that these expressions are
finite), and
\beq
K(x-y;3\la\phi_0^2)=16\pi^2
\int\frac{\rd^4 p}{(2\pi)^4}\int \frac{\rd^4 k}{(2\pi)^4}
\frac{e^{-\ri p(x-y)}}{(k^2+ 3\la\phi_0^2)((k+p)^2+ 3\la\phi_0^2)}
\eeq
Calculating the terms displayed in (\ref{gaexp}) (the relevant
integrals can be found in the
Appendix) we arrive at the following expression
\beq
\Ga^{(1)}(\phc)=\frac{9\la^2}{64\pi^2}\int\rd^4 x
\phc^2(x)\left[-\frac{1}{\eps}+\int_0^1\rd\al
\ln\left(\frac{-\al(1-\al)\Box+ 3\la\phi_0^2}{v^2}\right)
\right]\phc^2(x)
\label{Ganonlocal}
\eeq
The above result (\ref{Ganonlocal}) still contains the spurious
parameter $\phi_0^2$. Because the final result must be independent
of it, have definite conformal properties and at the same time
coincide with the Coleman-Weinberg potential in the limit of
constant $\phc(x)$, we conclude that, after including higher order terms
in (\ref{gaexp}) and (\ref{gaform}), the relevant part of
the renormalized non-local effective action should be obtained by the
replacement $\phi_0^2\rightarrow \phc^2(x)$. This gives
\beq
\Ga^{(1)}_R[\phc]=\frac{9\la^2}{64\pi^2}\int\rd^4 x
\,\phc^2(x)\left[\int_0^1\rd\al
\ln\left(\frac{-\al(1-\al)\Box+ 3\la\phc^2(x)}{v^2}\right)\right]\phc^2(x)
\label{Seff1}
\eeq
To be sure, one cannot hope to obtain a closed form analytical
expression for the complete one-loop effective action. Although
the above expression is certainly modified by further non-local
functionals at higher orders in the expansion, we claim that
(\ref{Seff1}) fully captures the {\em anomalous} conformal
properties of the one-loop effective action, in the sense that the
(unknown) functional modifications arising from convergent
contributions at higher orders will not affect the anomaly.
Observe that the result (\ref{Seff1}) cannot be expanded in $\la$
because the zeroth order term would be $\propto\ln(\Box)$ which is
ill-defined when acting on constant fields. Expanding in $\Box$,
on the other hand, the resulting series would involve inverse
powers of $\la$, and thus diverge at small $\la$.

While dimensional regularization automatically implements our
postulate of an exact conformal invariance of counterterm
dependent terms of the effective action, this is not so for other
regulators for which there appear non-conformal local terms at
intermediate steps of the calculation. For instance, with a
standard UV cutoff $\Lambda$, the divergent part of the effective
action (\ref{gaexp}) would read
\beq
\Ga^{(1)}[\phc]=\frac{3\la\La^2}{32\pi^2}\int\rd^4 x
\, \phc^2(x)-\frac{9\la^2}{64\pi^2}\ln\left(\frac{\La^2}{M^2}\right)
\int\rd^4 x \, \phc^4(x) \; + \; \dots
\label{gacut}
\eeq
It thus breaks conformal invariance, and furthermore depends
on the unphysical parameter $M^2$. Our basic requirement then amounts
to an {\em exact} cancellation of the first term by an appropriate
counterterm, not leaving any finite mass term either. Similarly, the
$M^2$-dependence of the second term is gotten rid of by renormalization.

\section{The conformal anomaly}

Exact conformal invariance is reflected in the conservation of the
conformal currents $J_\mu= \xi^\nu T_{\mu\nu}$ \cite{MS,CCJ}, with
the conformal Killing vectors $\xi_D^\mu = x^\mu$ for dilatations,
and $\xi^\mu_{K(\nu)} = 2 x^2 \delta^\mu_\nu - x^\mu x_\nu$ for
conformal boosts, where $T_{\mu\nu}$ is the conserved energy momentum
tensor; we have
\beq
\pa^\mu J_\mu = \pa^\mu ( \xi^\nu T_{\mu\nu})
    = \frac14(\partial_\nu \xi^\nu) \, T^\mu{}_\mu
\eeq
Current conservation thus implies $T^\mu{}_\mu =0$. If, on the
other hand, this symmetry is broken by quantum effects there will appear
an anomaly (trace anomaly) on the r.h.s. such that $T^\mu{}_\mu \neq 0$.
As is well known (see e.g. \cite{IZ,Weinberg}), anomalies may occur
when a symmetry of the classical Lagrangian cannot be maintained at the
quantum level. An anomaly is unavoidable when a symmetry breaking term
in the regulated effective action cannot be removed by a {\em local}
counterterm before the regulator is removed. This applies in particular
to conformal symmetry which is necessarily broken by {\em any} regularization,
see \cite{CDJ,F}.

The trace of the energy-momentum tensor can be calculated following
\cite{CCJ}, where it was shown that this trace vanishes for a conformally
invariant classical action if one makes use of the so-called improved
energy momentum tensor. The latter can be directly obtained by varying
the classical action (\ref{S}) w.r.t. the metric, which gives
\bea
T^{(0)}_{\mu\nu} &\equiv&
\frac{\delta S}{\delta g^{\mu\nu}}\Bigg|_{g_{\mu\nu}=\eta_{\mu\nu}} \nn\\
&&  \!\!\!\!\! \!\!\!\!\!\!\!\!\!\!\!\!\!\!\!\!\! = \;
\pa_\mu\phi\pa_\nu\phi-\eta_{\mu\nu}\left(
\frac12\pa_\al\phi\pa^\al\phi+\frac14 \la\phi^4\right)+
\frac16(\eta_{\mu\nu}\Box-\pa_\mu\pa_\nu)\phi^2
\eea
By contrast, the contribution to $T^{(1)}_{\mu\nu}$ calculated
from (\ref{Seff1}) has non-vanishing trace and constitutes the
conformal anomaly. The easiest way to calculate it is to use the
formula
\beq
T^\mu{}_\mu(x) =\left.\frac{\de\Ga[\phc,\om]}{\de\om(x)}\right|_{\om(x)=0}
\label{ttrace}
\eeq
where the functional derivative is calculated with respect to conformal
transformations (\ref{conftr}) and we put
$g_{\mu\nu}(x)=\eta_{\mu\nu}$ at the end. In order to derive this
from (\ref{Seff1}) we must thus properly covariantize all
expressions by re-inserting the metric $g_{\mu\nu}(x)$ in the
appropriate places.

To proceed further we rewrite (\ref{Seff1}) as
\bea\label{Ga2}
\Ga^{(1)}_R[\phc]&=&\frac{9\la^2}{64\pi^2}\int\rd^4 x \sqrt{g}
\,\phc^2(x) \times \\
&&\!\!\!\!\!\!\!\!\!\!\!\!\!\!\!\!\!\!\!\!\!\!\!\!\!\!\!\!\!\!\!\!\!\!\!\!
\left\{\ln\left(\frac{3\la\phi_c^2(x)}{v^2}\right) +\int_0^1\rd\al
\ln\left[ 1 + \frac{\al(1-\al)}{3\la} \frac{1}{\phi_c(x)}
\left(-\Box_g  +\frac16 R\right)\frac{1}{\phi_c(x)}\right]\right\} \phc^2(x)
\nn
\eea
Namely, using (\ref{conftr}) and (\ref{confinv}), it follows that
the second term inside parentheses is {\em invariant} under
conformal transformations: expanding the logarithm\footnote{As we
 said above, this should not be thought of as a proper perturbative
 expansion, although (\ref{Ga2}) may nevertheless serve as a possible
 definition of the logarithmic differential operator. The point
 here is simply to verify the proper behavior of this operator w.r.t.
 the Weyl scalings (\ref{conftr}).} we obtain an infinite series of terms,
each of which is scale-invariant due to the inverse powers of
$\phc(x)$. Let us mention here that the first order term in this
expansion produces a {\em finite} correction ({\it alias} wave function
renormalization) to the kinetic term in (\ref{S}). The first term under
the integral in (\ref{Ga2}) breaks conformal invariance and gives
the conformal anomaly
\beq
T^\mu{}_\mu(x)=\frac{9\la^2}{32\pi^2}\,\phc^4(x)
\label{anom}
\eeq
(on a curved space-time manifold this result is supplemented by the
well-known terms quadratic in the Riemann tensor). We have thus
confirmed (at this order) the general result that the anomaly
itself is a local expression, but is obtained as the functional
variation of a non-local expression (see e.g. \cite{MMM} for a
discussion of this point).

The above result can be easily generalized to the case of $O(N)$
symmetry, that is, to $N$ real scalar fields $\{ \phi_c^i (x) \,| \,
i=1,\dots,N \}$ transforming in the fundamental representation
of $O(N)$. The effective action is then equal to
\beq
\Ga^{(1)}[\phc]=-\frac{N-1}{2} \Tr\ln\left(1-\frac{\la\phc^2(x)}
{-\Box+\la\phc^2(x)}\right) -\frac12
\Tr\ln\left(1-\frac{3\la\phc^2(x)} {-\Box+3\la\phc^2(x)}\right)
\label{SeffON}
\eeq
where $\phc^2(x)\equiv \sum_i \phi_c^i(x)\phi_c^i(x)$, and the corresponding
conformal anomaly is
\beq
T^\mu{}_\mu(x)=\frac{(N+8)\la^2}{32\pi^2}\,\left(\phc^2(x)\right)^2
\label{anomON}
\eeq

These results can be rewritten in the form, familiar from general
discussions of the trace anomaly (see e.g. \cite{Duff,F}), viz.
\beq
T^\mu{}_\mu (x) = \beta (\lambda) O_4(x)
\label{Tbeta}
\eeq
where $O_4\propto \phc^4$ is a dimension four operator, and the
prefactor is the $\beta$-function for the $O(N)$ model (which is known 
up to five loops \cite{Kleinert})
\beq
\beta(\lambda)= \mu\frac{\pa\lambda}{\pa\mu}=\frac{N+8}{8\pi^2}\lambda^2 
\, + \, O(\lambda^3).
\eeq
Hence, at least to this order,
\beq
T^\mu{}_\mu (x) = \frac14\beta (\lambda) \left(\phc^2(x)\right)^2
\label{Tbetaexact}
\eeq
It is this relation, the {\em anomalous Ward identity}, which encapsulates 
the content of the symmetry, and how it is broken by quantum effects. A 
`good' regularization of the theory should therefore preserve the 
structure of the anomalous Ward identity (\ref{Tbeta}) as far as possible 
\cite{Bardeen}. From this point of view the advantages of dimensional 
regularization are evident: it preserves the structure of (\ref{Tbeta}) 
throughout the calculation by maintaining {\em exact} conformal symmetry 
at the level of the counterterms and the local part of the effective 
action, and in such a way that the parameter $M$ nowhere appears on the 
r.h.s. of (\ref{Tbeta}). Had we used a UV cutoff $\Lambda$ instead, there 
would have appeared spurious mass terms (depending on $M$ and $\Lambda$) 
on the r.h.s. of (\ref{Tbeta}) at intermediate steps of the calculation, 
as is evident from (\ref{gacut}). It is precisely the requirement of 
conformal invariance of the counterterm dependent terms of the effective 
action which ensures the absence of such spurious terms (as well as true 
mass terms) at any step of the calculation.

\section{The issue of quadratic divergences}

Let us now return to the issue of quadratic divergences in scalar
field theories. The calculation of the foregoing section shows
that the absence of (quadratically) divergent mass term
corrections can be consistently imposed order by order by
insisting on the {\em exact conformal invariance of the
counterterm dependent terms of the effective action}; the
effective CW potential then is the restriction of the renormalized
effective action to constant field configurations. This statement
is valid both in dimensional regularization (which does not
distinguish between quadratic and logarithmic divergences) as well
as in other schemes such as regularization in terms of an explicit
UV cutoff $\Lambda$. Therefore, the issue is not whether the
divergences which appear at intermediate steps of the calculation
are quadratic or not, but only whether or not a symmetry can be
imposed by means of local counterterms, and how the remaining
anomalous terms can be uniquely characterized and computed. All
this is, of course, in complete accord with standard
renormalization theory
\cite{IZ,Weinberg}.

To see that the way by which (classical) conformal invariance disposes
of quadratic divergences is really no different from the way in which
supersymmetry takes care of the problem, it is useful to recall that there
is no problem whatsoever (other than convenience) in regularizing a
supersymmetric theory by means of a {\em non}-supersymmetric regulator
\cite{Maison} --- such as, for instance, ordinary dimensional regularization,
or the use of {\em different} cutoffs for bosonic and fermionic loops.
In both cases supersymmetry is violated but can be re-instated order by
order by means of appropriate counterterms~\footnote{This is true because
 there do exist perturbative regulators preserving supersymmetry, such
 as higher derivative regulators \cite{IlZu}, dimensional regularization
 by dimensional reduction \cite{Siegel} (whose status at higher loops
 remains uncertain, however), or superspace methods (see e.g.\cite{FE}).}
(which themselves then also violate supersymmetry). In other
words, the celebrated cancellation of quadratic divergences in
supersymmetric theories is thus by no means automatic, but the
result of an order by order imposition of supersymmetry at the
level of the counterterm dependent terms of the effective action
in perturbation theory. Furthermore, as emphasized in
\cite{Bardeen}, quadratic divergences have no import on the general
structure of the anomalous Ward identity (\ref{Tbeta}) because the
one-loop $\beta$-function on the r.h.s. of (\ref{Tbeta}) `does not know'
about them, and therefore their appearance should be rather viewed
as an artifact of the particular method employed to regulate the theory.

While the perturbative renormalization procedure thus does not care as
to whether the divergences, which appear at intermediate steps of the
calculation, are logarithmic or quadratic, the picture is, however,
different in a Wilsonian perspective. There, one views the SM as
being embedded as an effective low energy theory in some more unified
theory whose modes above the weak scale have been `integrated out'.
Because one would then expect the mass corrections to be of the order
of the unification scale (which would act as an effective UV cutoff),
the existence of nearly massless modes (in comparison with the unification
scale) indeed becomes a problem in the absence of an independent
reason for their existence (as we pointed out already, conformal
invariance cannot be invoked in the presence of a large mass scale).
This conclusion seems inevitable if the
large scale theory is also described by quantum field theory -- as is
the case for most `beyond the SM' scenarios, such as GUTs and the MSSM.
For this reason, and as already stated in the introduction, {\em the scenario
proposed in \cite{MN} can only work if there are no intermediate mass
scales of any kind between the weak scale and the Planck scale}.
In addition we must require that the running couplings stay positive 
and bounded over a large interval of energies from $\Lambda_{QCD}$ up 
to the Planck scale, since otherwise the theory would break down in between.
The QCD scale is a natural IR cutoff because below that scale the 
conformal symmetry is known to be broken by nonperturbative QCD effects 
(quark and gluon condensates) which introduce their own mass scale. The 
upper limit derives from our assumption that physics at and beyond the Planck 
scale is no longer described in terms of standard quantum field theory, see 
the comments below. For this reason neither possible IR fixed points
nor UV fixed points are relevant to the present considerations.   

Let us explain why, in our opinion, the Wilsonian arguments outlined
above may not be applicable to the Planck scale $M_P$, the only large 
scale in nature of whose existence we can be sure. Namely the usual 
arguments leading to divergent loop integrals cut off at $M_P$ are based 
on the tacit assumption that ordinary quantum field theory works smoothly 
all the way up to $M_P$, then to be abruptly replaced by a Planck scale
theory of quantum gravity. However, we would expect that such a theory
which (by some as yet unknown mechanism) is supposed to give rise to an
effectively conformal theory below the Planck scale, itself can {\em not}
be a space-time based quantum field theory. Rather, space-time and its
concomitant symmetries would then be emergent properties in a theory of
quantum gravity~\footnote{Indeed, recent investigations on infinite
 dimensional hidden symmetries in supergravity suggest precisely such
 a scenario, see e.g. \cite{DN}.}, as would be the case for quantum field
theory. Even if field theory methods were to apply right up to $M_P$,
such arguments do not take into account the anticipated UV finiteness
of a proper (unified) theory of quantum gravity, and the fact that the
resulting UV cancellations at the Planck scale may survive to low energy
scales with conformal symmetry and in the absence of intermediate scales.

We note that several of the points raised here were also made
in a recent preprint \cite{Shap}. The main difference is that the
$\nu$MSM model proposed there breaks conformal invariance already
at the classical level, because the extra scalar field there is
supposed to play the role of the inflaton. This requires not only
a special fine-tuning of the parameters, but in particular, an
explicit mass term at variance with conformal invariance; the latter
is needed because the CW mechanism is not compatible with the values
of the scalar self-couplings required to reproduce inflation. By contrast,
we here make no attempt to use scalar fields for such purposes; rather, it
is supposed that the mechanism leading to inflation -- or what effectively
{\em looks like it} from our low energy vantage point -- is intrinsically
quantum gravitational in nature. Interestingly, the scenarios of \cite{MN}
and \cite{Shap} differ in their predictions for the Higgs mass spectrum,
and may thus be subject to experimental discrimination (and falsification).

To conclude: it is often said that the worst case for high energy
physics would be if LHC discovered only the Higgs particle(s), but
nothing else. We think otherwise: if there are no intermediate scales
there is nothing to obstruct our view of the Planck scale. The challenge
would then be to explain the observed structure of low energy physics
directly and in a minimalistic way from a Planck scale theory of quantum
gravity and quantum space-time, rather than evade the problem by
introducing myriads of new particles and couplings, whose direct
verification may well remain out of experimental reach. Besides, when
trying to solve the hierarchy problem one is {\it a priori} in a much
better position if the only terms in the effective action which break
conformal invariance are logarithms.

\section{Appendix}

For the convenience of the reader, we here collect some integrals
used in the main body of this paper. For the dimensionally
regulated integrals we have
\bea
\int\frac{\rd^d k}{(2\pi)^d}
\ln\left(1+\frac{M^2}{k^2}\right)
&=&\frac{4\Ga(2-d/2)M^d}{(4\pi)^{d/2}d(2-d)}\\
\int\frac{\rd^d k}{(2\pi)^d}
\frac{1}{k^2+M^2}
&=&\frac{2\Ga(2-d/2)M^{d-2}}{(4\pi)^{d/2}(2-d)}\nn\\
\int\frac{\rd^d k}{(2\pi)^d}
\frac{1}{(k^2+M^2)((k+p)^2+m^2)}
&=&\frac{\Ga(2-d/2)}{(4\pi)^{d/2}}
\int_0^1\frac{ \rd\al}{ (\al(1-\al)p^2+M^2)^{2-d/2}}\nn
\eea
For $d=4-2\eps$ and $C=e^{\ga}/(4\pi)$ it gives up to $O(\eps)$
\bea
16\pi^2(Cv^2)^\eps\int\frac{\rd^d k}{(2\pi)^d}
\ln\left(1+\frac{M^2}{k^2}\right)
&=&-\frac{M^4}{2\eps}+\frac{M^4}{2}\left(\ln\frac{M^2}{v^2}
-\frac32\right) \nn\\
16\pi^2\int\frac{\rd^d k}{(2\pi)^d} \frac{(Cv^2)^\eps}{k^2+M^2}
&=&-\frac{M^2}{\eps}
+M^2\left(\ln\frac{M^2}{v^2}-\frac12\right) \; \nn\\
\int\frac{\rd^d k}{(2\pi)^d}
\frac{16\pi^2(Cv^2)^\eps}{(k^2+M^2)((k+p)^2+M^2)}
&=&\frac{1}{\eps}-\int_0^1 \rd\al
\ln\left(\frac{\al(1-\al)p^2+M^2}{v^2}\right)
 \nn
\eea

\vspace{4.0mm}
\noindent
{\bf Acknowledgments:} We are grateful to Luis Alvarez-Gaum\'e 
and Misha Shaposhnikov for discussions and correspondence, 
and to the referee for comments. K.A.M was partially 
supported by the EU grant MRTN-CT-2006-035863 and the Polish grant  
N202 081 32/1844. K.A.M. thanks the Albert-Einstein-Institut for
hospitality and financial support.

\end{document}